\documentclass[11pt]{article}

\usepackage[utf8]{inputenc}
\usepackage{amsfonts,amssymb,amsmath,enumerate}
\usepackage{theorem,cite}
\usepackage{indentfirst}
\usepackage{textcomp}
\usepackage{color}
\usepackage{varioref}
\usepackage{float}
\usepackage{array}
\usepackage[nice]{nicefrac}
\usepackage{caption}
\usepackage{booktabs}
\usepackage{multirow} 
\usepackage{latexsym}   
\usepackage[colorlinks=true]{hyperref}
\usepackage{epsfig}
\usepackage{pgf,tikz}

\definecolor{dblue}{rgb}{0,0,0.7}
\hypersetup{colorlinks,citecolor=dblue,filecolor=black,linkcolor=dblue,urlcolor=blue}
\definecolor{Aqua}{RGB}{100, 0, 100} 
 
\definecolor{dog}{RGB}{0,153 0}

\numberwithin{equation}{section}

\begin{document}
\title{\vspace{-6ex}\bf Generalized quadratic commutator algebras of PBW-type \vspace{-1ex} }
\author{Ian Marquette$^1$\footnote{i.marquette@uq.edu.au}, Luke Yates$^2$\footnote{luke.yates@utas.edu.au}$\,$and Peter Jarvis$^2$\footnote{peter.jarvis@utas.edu.au}
\\[1cm]
$^1$ School of Mathematics and Physics, The University of Queensland,\\ St Lucia, Brisbane,
Queensland, 4072, AUSTRALIA\\
$^2$ School of Natural Sciences, University of Tasmania,\\ Hobart, Tasmania, Australia}
\maketitle

\abstract

In recent years, various nonlinear algebraic structures have been obtained in the context of quantum systems as symmetry algebras, Painlev\'{e} transcendent models and missing label problems. In this paper we treat all of these algebras as instances of the class of quadratic (and higher degree) commutator bracket algebras of PBW type. We provide a general approach for simplifying the constraints arising from the diamond lemma, and apply this in particular to give a comprehensive analysis of the quadratic case. We present new examples of quadratic algebras, which admit a cubic Casimir invariant. The connection with other approaches such as Gr\"{o}bner bases is developed, and we suggest how our explicit and computational techniques can be relevant in other contexts.
\vspace{5mm}


\tableofcontents


\section{Introduction}
\label{sec:Introduction}

\subsection{Background}
\label{subsec:Background}
The role of symmetry in  physical systems has long been recognised since the seminal work of Lie, on groups of continuous transformations, the associated mathematical framework of Lie groups and Lie algebras, and the crucial role of conservation laws and the theorems of Noether, in the characterization of the underlying dynamics. Thus, application of symmetry principles and associated conserved quantities has a natural relationship with commutator bracket type algebraic relations, and commutative subalgebras. Refining the analysis to the interplay of bosonic and fermionic parts, the graded extension of these considerations leads further to the inclusion of anticommutator-type algebras, in the mathematical framework of Lie superalgebras. 

In quantum physics in the study of superintegrable systems, it has been observed as a common feature that there also exist underlying quadratic \cite{gra92,das01,post11,mil13b} , cubic \cite{mar09} and higher order polynomial algebras \cite{mar14,bas19,mar20,cram20}: that is, where the the exchange relations of the generators (commonly, of the form of commutator-type algebraic relations) are of degree quadratic or higher. Higher order symmetries can be interpreted in the context of Lie symmetries and the generalised symmetries \cite{sh01,mil84}. Usually the generators of those algebraic structures are quantum mechanical operators which are polynomial in momenta and commute with the Hamiltonian.

In such studies, the polynomial algebras typically involve explicit differential-operator, oscillator or orthogonal-polynomial realizations. This close connection with different types of special functions and orthogonal polynomials is one of the main technical motivations for the recent growing body of work on the associated algebraic structures, in particular on their generalisation to the special class so-called higher rank quadratic algebras \cite{gab18,bie20,Lati21}. In terms of applications, these algebraic structures have been used to get insight in a range of areas: in quantum systems, but also in other fields of applied mathematics such as band time limiting \cite{alb18}. Amongst the properties of the polynomial algebras carried over from, and familiar from standard symmetry algebras, is that of associativity (or, in terms of the commutator-bracket-type relations, of satisfying the Jacobi identity), guaranteed in this context by the fact they are constructed via explicit realizations of the generators in terms of an underlying associative algebra. Also retained in general is the existence of Casimir invariants, which continue to play a central role.

Fulfillment of the Jacobi identity, or its generalisation to graded and/or non-bracket-type algebras, guarantees the existence of a Poincar\'e-Birkhoff-Witt (PBW) basis: an ordered monomial basis expressed in terms of the generators of the associative algebra. At a practical level, the availability of a PBW basis facilitates computation and comparison of polynomial expressions and their products; for example, to verify (or discover) the existence of Casimir invariants \cite{mar09,mar14}.

The purpose of this paper is to provide a systematic method to generate new families of polynomial commutator-type PBW algebras for a given maximal polynomial degree of the exchange relations. The constraints imposed by the (generalised) Jacobi identities are difficult to deal with, even for Lie algebras, and in the general context of polynomial algebras they lead to complicated systems of algebraic equations for which we provide solution strategies. 

\subsection{A working example: the Daskaloyannis algebra}
\label{subsec:Daskaloyannis}
As an example of the construction of polynomial algebras in this genre, we outline here the well-known Daskaloyannis algebra \cite{das01}, which also includes the original three-dimensional quadratic Racah algebra \cite{gra92}. As will emerge from our treatment (see below), this algebra is, in turn, a special case of a more general family of such three-generator quadratic algebras. We take the commutation relations of the generators $A, B, C$ to be
\begin{eqnarray}
\label{eq:daskaloyannisquadratic}
     {[}A,B{]}&=& C \nonumber \\
    {[}A,C{]} &=& \alpha A^2 + \beta \{A,B\} + \gamma A + \delta B + \epsilon  \\
    {[B},C{]} &=& \nu A^2  - \alpha \{A,B\} - \beta B^2 + \xi A - \gamma B + \zeta \nonumber
\end{eqnarray}
The case $\nu=0$ corresponds to the Racah algebra which, with its generalisations to higher rank algebras, has received of lot of attention in the recent literature \cite{Lati21,Latini2021b,Kuru2020,Huang2020,DeBie2021}. The (abstract) algebra defined by the relations \eqref{eq:daskaloyannisquadratic} has an ordered PBW basis (due to its satisfaction of the generalised Jacobi identities) and possesses the following cubic Casimir operator (see section \ref{subsec:theoryGradedPBW} for definitions):
\begin{equation}
\label{eq:daskCasimir}
\begin{split}
K = C^2 - \alpha \{A^2,B\} - \beta \{A,B^2\} + ( \alpha \beta - \gamma) \{A,B\} + ( \beta^2 - \delta) B^2 +\\
 \quad ( \beta \gamma - 2 \zeta) B + \frac{2a}{3} A^3 + ( \xi + \frac{\nu\beta}{3} + \alpha^2 ) A^2 + ( \frac{\nu\delta}{3} + \alpha \gamma +2 \zeta ) A
 \end{split}
\end{equation}
The original algebra considered by Daskaloyannis differed slightly from the parametric relations \eqref{eq:daskaloyannisquadratic}, as $\alpha$, $\beta$ and $\mu$ were held constant, however the fulfilment of commutation relations and of the Jacobi identities required the other structure constants to depend on a central element $H$ (a Hamiltonian, in the context of superintegrable systems). The additional terms  $\nu\neq0$ allowed the algebra to be applied to wider classes of quantum systems, in particular position dependent models \cite{Quesne2007a,Quesne2007b}, which take into account different effects such as curvature or deformation of canonical commutation relations, and can apply in contexts such as semiconductors and quantum liquids. The specific functional dependence of the structure constants on $H$ derives from the coefficients of an associated generalized oscillator construction \cite{mar09,mar14}, used in the context of quantum systems to construct energy eigenstates. In this realization, the Casimir \eqref{eq:daskCasimir} can be expressed as a polynomial in $H$. Such a solution is constructed for a generalized two-dimensional Kepler problem in \cite{das01}.

\subsection{Summary and main results}
\label{subsec:Summary}
As is evident from this quadratic example, investigations of polynomial algebras frequently derive from scenarios in which there is an associated explicit realization, either by way of a generalized Fock space, as above, or equivalent differential operator transcriptions; moreover they are characterized by a high degree of flexibility in the parametric dependence of their structure constants. This property also persists at higher degree. For instances of cubic algebras, arising from the Lie algebra $sl(3)$\, and its enveloping algebra, we cite \cite{vin20} (missing label problem)\,, \cite{flat89} (decomposition of the enveloping algebra), and \cite{cor21} (commutant structure). Furthermore, in \cite{cor21,vin20}\,, mappings from the cubic algebras back to related quadratic algebras are studied; in such cases, structure constants typically become constrained as functions of certain central elements. Related examples include the deformation of Sklyanin algebras \cite{sat20} and Lotka-Voltera models with quadratic Poisson structure \cite{kamp14}; however, quantization of the latter remains to be studied. Beyond this, in \cite{yat18, PDJRudolphYates2011,ragoucy2020finite} structural considerations driven by the necessity to resolve ambiguities in establishing a PBW basis, lead to the identification of certain types of quadratic superalgebra (see also \cite{mar09,vin20}). Another type of relation involving the so-called $q$-commutator was introduced \cite{Terwilliger2005} in the context of tridiagonal operators: now known as the Leonard pairs, which has attracted some attention with a connection to $q$-Racah polynomials \cite{Nomura2021, Terwilliger2004}.

In spite of such wide-ranging case studies, and the interrelations which have been established between Lie algebras and quadratic algebras, and quadratic and cubic algebras, there is as yet no analogue of a classification of low-dimensional polynomial algebras.
In this work, we propose a concrete and systematic approach for the construction of certain types of polynomial algebras. We go beyond the context of integrable systems, and related algebraic structure provided by explicit differential operator or Fock space realizations, as a first step towards classifying polynomial algebras; in particular, the simplest quadratic and cubic algebras involving three generators, from a purely algebraic standpoint. We will rely on a graded-degree construction and certain transformations which preserve the PBW basis \cite{li14,vin20}. Classes of algebraic structures informing such examples, for example premised on Koszulness or Gr\"{o}bner basis formalisms, are treated more abstractly in \cite{pol05,berg78,li14}.

The paper is organised in the following way. In Section 2, we present a review of the relevant background theory and introduce the main definitions and notational conventions. Using the notion of a graded-degree function in conjunction with the diamond lemma, we develop a methodology for the obtaining the most general set of solutions of the PBW constraints, generating parametrised families of polynomial PBW-algebras for a given maximum polynomial degree and fixed number of generators. As a case study, we specialise to cubic algebras with three generators to generalise the family of Calabi-Yau algebras. In Section 3, we give a detailed illustration of the method applied to quadratic algebras with three generators. Our computational strategy relies on a lower-triangular basis transformation which leads to a characterisation of the solution set according to the structure of admissible canonical forms of the defining relation associated with `lowest'-ordered bracket (i.e., $[A,B]$, where $A<B<C$). The type is determined by the inclusion of the generator $C$ (i.e., the existence [type 1] or non-existence [type 2] of a two-dimensional subalgebra generated by $A$ and $B$) and further sub-types (i.e., (1a), (1b), (1c), (1d), (2a), (2b), \ldots, (2h)) are determined by the included combination of the remaining eligible monomial terms. Solutions can be further grouped according to existence of a cubic Casimir operator, whose explicit form (when it exists) can be discovered in a constructive manner using the PBW basis. The subclass (1a) possessing a cubic Casimir generalises the examples obtained in \cite{mar09,cram20}. In Section 4, we formulate the notion of a Casimir invariant operator for polynomial PBW-algebras, and present explicit constructions of Casimir invariants of cubic degree, for a selection of 
subclasses of standard forms based on the leading $[A,B]$ bracket structure. In the Appendix, we augment the typology of canonical forms and solutions by including the possibility of central terms, and we present an example with one central element.

\section{Review of theory}
\label{sec:theory}

In the first part of this section, we recall existing theoretical frameworks to define polynomial algebras from a set of defining relations. Using a monomial ordering based on graded degree, we obtain a canonical spanning set of monomials which serve as a candidate PBW basis for a given set of relations. We define a class of commutator-type algebras and use the diamond lemma to derive conditions under which the canonical monomials are a PBW basis. These conditions yield constraints on the structure constants---the coefficients of monomial terms in the defining relations---and generalise the usual Jacobi identities from Lie theory. In the second part, we provide a working example of this framework, using the graded-degree structure to incrementally generalise the quadratic Daskaloyannis algebra \eqref{eq:daskaloyannisquadratic} to a new parameteric family of cubic extensions: a further generalisation of existing cubic extensions and the cubic Calabi-Yau algebra.

\subsection{Graded-degree functions and algebras of PBW type }
\label{subsec:theoryGradedPBW}

Given a finite $\mathbb{K}$-vector space $X$  (here $\mathbb{K}= \mathbb{R}\,\mathrm{or}\,\mathbb{C}$), the defining (or structural) relations of a polynomial algebra are a set of polynomial expressions in the generators of $X$ that define the multiplication rules of the algebra. These expressions can be viewed as a generating set for a two-sided ideal within the corresponding tensor algebra $T(X)$. Consider the broadly encompassing class of algebraic structures whose defining relations are of the following general form:
\begin{equation}\label{eq:R_general}
    R\subset \mathbb{K} + X + (X\otimes X) + (X\otimes X\otimes X) + \cdots,
\end{equation}
Associated with $R$ is the (non-commutative) polynomial algebra, obtained via the factorisation
\begin{equation}
    U = T(X)/\langle R \rangle
\end{equation}
where $\langle R \rangle$ denotes the two-sided ideal generated by $R$ in $T(X)$. 

To explore the structure of $U$, it is useful to define a graded-lexicographic monomial ordering. Take $x_i$, $i=1,2,...,n$ as a basis for $X$. A \emph{graded-degree} function $d: X\rightarrow \mathbb{N}$ assigns to each basis element a positive integer $d(x_i)=m_i$. The definition of the graded degree can be extended to both a monomial term $a = x_{i_1}\!\cdots x_{i_s}$ given by $d(a)=m_{i_1}+\cdots+m_{i_s}$ and to a polynomial expression $f = \sum_{i=1}^t \lambda_i a_i $, $\lambda_i\in\mathbb{K}$, given by $d(f)=\mathrm{max}\{d(a_i)\mid i = 1,...,t\}$.  (Throughout, we adopt the shorthand notation $x_ix_j\equiv x_i\otimes x_j$ etc.) Following \cite{Li2011}, the graded-degree function can be used to define the graded-lexicographic monomial ordering
\begin{eqnarray}\label{eq:grdeglex}
a \prec b \Leftrightarrow \left\{
\begin{array}{ll}
d(a)<d(b)\\
d(a)=d(b) \mbox{ and } a \prec_{{}_\mathrm{lex}} b
\end{array}  \right.
\end{eqnarray}
where $\prec_{{}_\mathrm{lex}}$ is the usual lexicographic ordering based on the principle of first difference with respect to the ordering of the basis elements of $X$. The most commonly used graded-lexicographic monomial ordering is the degree-lexicographic ordering which corresponds to the case $d(x_i)=1$, $i=1,...,n$.

Each defining relation $r\in R$ admits a polynomial representation for which the leading monomial $\mathrm{LM}(r)$ is defined as the largest monomial with respect to the ordering \eqref{eq:grdeglex}. We set $\mathrm{LM}(R)=\{\mathrm{LM}(r)\mid r\in R\}$. The identification of a leading term transforms each defining relation into a corresponding substitution rule: $\mathrm{LM}(r) \mapsto r - \mathrm{LM}(r)$, which allows an arbitrary $U$-monomial to be reduced to a sum of minimal (or irreducible) terms. For example, given the ordered basis $x_1,x_2,x_3$ for $X$, with $d(x_1)\!=\!d(x_2)\!=\!d(x_3) = 1$, the $U$-element $x_2x_1$ can be reduced to $x_1x_2 + x_3$ if the relation $x_1x_2-x_2x_1 + x_3$ belongs to $R$. A monomial $u=x_{i_1}x_{i_2}\cdots x_{i_s}$ is called \emph{reducible} if it possible to reduce its order by applying a substitution rule; that is, if there exists $j$ such that $x_{i_j}x_{i_{j+1}}\in\mathrm{LM}(R)$. Otherwise, it is \emph{irreducible}. 

A \emph{PBW-algebra} is a polynomial algebra $U$ for which $\mathrm{LM}(R)\subset X\otimes X$ and the monomial elements
\begin{equation}\label{eq:pbw_priddy}
    \{x_{k_1}\cdots x_{k_K}\mid x_{k_j}x_{k_{j+1}}\notin \mathrm{LM}(R),\, j = 1,...,K-1, K\in\mathbb{N}\}
\end{equation}
are a basis for $U$, called a PBW basis. Despite the constraint that the leading monomial is quadratic, the freedom to set different graded-degree values for each generator allows higher order polynomial terms to be included. The existence of a PBW basis depends on both the ordering of the basis elements of $X$ and the precise form of the structural relations \eqref{eq:R_general}. An important class of PBW-algebras are those for which the PBW basis is the set of ordered monomials 
\begin{equation}\label{eq:x_alpha}
    \{x^{\alpha}=x_1^{\alpha_1}x_2^{\alpha_2}\cdots x_n^{\alpha_n}\mid \alpha = (\alpha_1,...,\alpha_n), \alpha_k\in \mathbb{N}\}
\end{equation}
%
Intuitively, a prerequisite for the monomials \eqref{eq:x_alpha} to be a spanning set for $U$ is the existence of a substitution rule for each of the $n(n-1)/2$ unordered monomial pairs such that $\mathrm{LM}(R) = \{x_jx_i\mid i<j\}$. Taking a constructive approach to the characterisation of PBW-algebras within this class, we consider commutator algebras with relations of the general form
\begin{eqnarray}\label{eq:Rji}
    R = \{r_{ji} = x_jx_i -\lambda_{ji}\,x_ix_j - p_{ji}\mid i<j,\,i,j=1,...,n\}\qquad
\end{eqnarray}
where $\mathrm{LM}(r_{ij}) = x_jx_i$, $\lambda_{ji}\in\mathbb{K}$, and 
$p_{ij} = \sum_\alpha c^\alpha_{ij}x^\alpha$, $c^\alpha_{ij}\in\mathbb{K}$.

By construction, the monomials \eqref{eq:x_alpha} (or indeed \eqref{eq:pbw_priddy} more generally) are a spanning set for $U$, but the requirement of their linear independence imposes further constraints on the structure constants. The problem of determining their linear independence is equivalent to the resolution of the reduction ambiguites that arise when applying the substitution rules. For example, the unordered monomial $x_3x_2x_1$ can be reduced in two ways: one is to begin the reduction from the right-hand-side by substituting $x_2x_1$ with $r_{21} - x_2x_1$ and the other is to first reduce from the left-hand side substituting for $x_3x_2$, in both cases applying further reductions until the resulting polynomial is irreducible. The ambiguity concerning the choice of reduction sequence is said to be resolved if the same irreducible polynomial is obtained regardless of the sequence. In other words, the monomials $\eqref{eq:pbw_priddy}$ are a basis for $U$ if all reduction ambiguities are resolved. This idea is formalised in the diamond lemma \cite[Theorem 1.2]{berg78}, where, provided that $\prec$ is a partial order, satisfying $b\prec b' \Rightarrow abc\prec ab'c$ ($a,b,b',c \in T(X)$), induction on the monomials as ordered by $\prec$ reduces the general problem of resolution of all ambiguities to resolution of all ambiguous triples. In the present setting, where $\mathrm{LM}(R)\subset X\otimes X$, the ambiguous triples are the (polynomial) degree-three monomials belonging to the set $S = \{x_kx_jx_i\mid x_kx_j\mbox{ and } x_jx_i \in \mathrm{LM}(R)\}$.

To illustrate the role of the graded-degree function and the application of the diamond lemma, we consider the general three-generator commutator algebras with generators $A$, $B$, $C$, and ordering relations:
\begin{eqnarray}
    &\,A\prec_{\mathrm{lex}} B \prec_{\mathrm{lex}} C \qquad\qquad\ &\mbox{(lexicographic order)}\\
    &d(A) \leq d(B) \leq d(C) \qquad &\mbox{(graded degree)}
\end{eqnarray}
setting (without loss of generality) $d(A) = 1 + a$, $d(B)=d(A) + b$, and $d(C) = d(B) + c$, for fixed $a,b,c\in\mathbb{N}$.
From \eqref{eq:Rji}, there are three defining relations $r_{ji}$ for which monomial terms within the corresponding polynomial expressions $p_{ji}$ are constrained in (polynomial) degree by the choice of $a$, $b$, and $c$ such that $d(p_{ji})<d(x_jx_i)$. Expressing the $p_{ji}$ schematically (i.e., without showing the structure constants) the explicit dependence of the inclusion of candidate monomial terms on the choice of graded-degree values can be written in terms of inequalities:
\begin{equation}\label{eq:rhsGradedDegTerms}
\begin{aligned}
    p_{BA} ={}& A + B & \\[-1mm]
    & \,\,+C &{}(a+b > c)\\[-1mm] 
    & \,\,+A^2 &\\[-1mm]  
    & \,\,+A^3 &{}(a\leq b-1) \\[-1mm]  
    & \,\,+\cdots & (\cdots) \\[5mm] 
    p_{CA} ={}& A + B + C &\\[-1mm] 
    & \,\,+A^2 + AB + AC + B^2& \\[-1mm] 
    & \,\,+A^3 &{}(a\leq b + c-1) \\[-1mm] 
    & \,\,+A^2B &{}(a\leq c-1) \\[-1mm] 
    & \,\,+AB^2 &{}(a+b\leq c-1) \\[-1mm] 
    & \,\,+B^3 &{}(a+2b\leq c-1) \\[-1mm] 
    & \,\,+ \cdots& (\cdots)\\[5mm] 
    p_{CB}={}& A + B + C &\\[-1mm] 
    & \,\,+A^2 + AB + AC + B^2 &\\[-1mm] 
    & \,\,+A^3 &(a\leq 2b + c-1)\\[-1mm]  
    & \,\,+A^2B &(a\leq b + c-1)\\[-1mm]  
    & \,\,+A^2C &(a\leq b -1) \\[-1mm]  
    & \,\,+AB^2 &(a\leq c-1) \\[-1mm]  
    & \,\,+B^3 &(a+b\leq c-1) \\[-1mm]  
    & \,\,+\cdots & (\cdots) 
\end{aligned}
\end{equation}
Since the induction procedure which underpins the diamond lemma relies only on the existence of a ordering relation $\prec$, its precise form is inconsequential. Thus, in the current setting, it is sufficient to determine the existence of degree-function values which satisfy the corresponding inequality for each included monomial term. For the three-generator case, the set $S$ of unordered degree-three monomials contains just one element: $CBA$. Applying the diamond lemma, we reduce $CBA$ from both the left and right, setting the difference of the two reductions to zero to identify the constraints for $U$ to be of PBW-type, explicitly:
\begin{eqnarray*}
CBA &\xrightarrow[]{\mathrm{LHS}}&(\lambda_{CB} BC-p_{CB})A\\
&=& \lambda_{CB} B(\lambda_{CA}AC- p_{CA}) - p_{CB}A\\
&=& \lambda_{CB}\lambda_{CA}(\lambda_{BA}AB- p_{BA})C - \lambda_{CB}B p_{CA}- p_{CB}A\\
&=& \lambda_{CB}\lambda_{CA}\lambda_{BA}ABC - \lambda_{CB}\lambda_{CA} p_{BA}C - \lambda_{CB} B p_{CA}- p_{CB}A\qquad\qquad(\mbox{L})\\[5mm]
CBA &\xrightarrow[]{\mathrm{RHS}}&C(\lambda_{BA}AB- p_{BA})\\
&=& \lambda_{BA}(\lambda_{CA}AC- p_{CA})B - C p_{BA}\\
&=& \lambda_{BA}\lambda_{CA}A(\lambda_{CB} BC-p_{CB}) - \lambda_{CA}p_{BA}B - Cp_{BA}\\
&=& \lambda_{BA}\lambda_{CA}\lambda_{CB}ABC -  \lambda_{BA}\lambda_{CA}Ap_{CB}- \lambda_{CA}p_{BA}B - Cp_{BA} \qquad\qquad(\mbox{R})
\end{eqnarray*}
Setting $(\mbox{L})-(\mbox{R})$ to zero gives
\begin{align}\label{eq:diamond_CBA}
0=\lambda_{CB}\lambda_{CA} p_{BA}C - Cp_{BA} + 
\lambda_{CB} B p_{CA} - \lambda_{CA}p_{BA}B +
p_{CB}A - \lambda_{BA}\lambda_{CA}Ap_{CB}
\end{align}
The constraints \eqref{eq:diamond_CBA} are a generalisation of the well-known Jacobi identities for Lie algebras (i.e., $\lambda_{ji}=1$ and $p_{ji} = [x_i,x_j] =  \sum_{k=1}^n c_{ji}^k x_k$). 

The diamond lemma provides a practical means to test for the existence of a PBW basis, but other equivalent approaches have been developed in different contexts (see \cite{Shepler2014} for a review). For quadratic algebras, homological methods have been used to derive the conditions in which well-understood homogeneous algebras (e.g., the symmetric algebra) are isomorphic to associated graded versions of their deformed non-homogeneous counterparts (e.g., Lie algebras and various generalisations) \cite{pol05,Braverman1996}. In Gr\"obner basis theory, common algorithms to compute Gr\"obner bases (the generating set of an ideal in a polynomial ring) are based on polynomial reductions and the resolution of (overlap) ambiguities \cite{Li2011,Buchberger1985}. The practical equivalence of these approaches emerges from the similarity of the calculations that are ultimately required when applying them to specific algebras. Indeed, elements of the diamond lemma were nascent in the original proofs of the PBW theorem for the universal enveloping algebra of a Lie algebra given by Poincar\'e, Birkhoff and Witt \cite{Shepler2014}.

\subsection{A case study: cubic deformations and the Calabi-Yau algebra}
\label{subsec:Deformation}
In this subsection we provide, as a further working example, generalizations of the quadratic Daskaloyannis \cite{das01} three-generator commutator-type algebra 
(see section\ref{subsec:Daskaloyannis} above) to polynomial degree three and beyond. We show how these cubic cases can be systematically recovered from the quadratic case, consistently with the graded degree formalism, by incrementally modifying the bracket relations by the addition of an extra term or terms.

The following cubic extension of the Daskaloyannis quadratic algebra was first derived \cite{mar09} via generalized oscillator constructions. The defining relations, expressed as commutator bracket relations in $U$ (i.e., $[x_j,x_i] = x_jx_i-x_ix_j = p_{ji}$, c.f. \eqref{eq:Rji} with $\lambda_{ji}=1$), for the generators $A, B, C$ are 
\begin{align}\label{eq:cubic}
	&[B,A] = C\nonumber\,,\\
	&[C,A] = \alpha A^2 + 2\beta AB + \gamma A + \delta B + \beta C + \epsilon\,, \\
	&[C,B] = \mu A^3 + \nu A^2  - 2\alpha AB  - \beta B^2  + 
	\xi A - \gamma B -\alpha C + \zeta\,, \nonumber
\end{align}
expressed in terms of structure constants given by free parameters 
$\alpha, \beta, \gamma, \delta, \mu, \nu, \xi$\,, central generators associated with $\epsilon, \zeta$\,, and with all generalized Jacobi identities satisfied. The Daskaloyannis quadratic algebra (\ref{eq:daskaloyannisquadratic}) is the case $\mu=0$\,. 

A further extended form of this algebra was subsequently recovered in the context of the $sl(3) \times sl(3)$ Clebsch-Gordon series and integrable systems with third order integrals of motion, as the so-called  ``Calabi-Yau algebra'' (\hspace{1sp}\cite{vin20}; for details see below). Other examples with cubic deformations occur for Heun \cite{bas19}, Heun-Lie \cite{alb18}, and Askey-Wilson-Heun \cite{sat20} algebras, with realizations obtained via differential operators or via Lie algebra structure \cite{flat89,cor21}\,. 

Given our discussion in section \ref{subsec:theoryGradedPBW} above, and the structure of the Daskaloyannis algebra (\ref{eq:daskaloyannisquadratic}), and its cubic extension (\ref{eq:cubic}) above\,,  we assume there is a graded degree assignment of the type  discussed in the three-generator case. In the notation (\ref{eq:grdeglex}) above, given $a<b<c$\,, as noted, the existence of the bracket relation ${[}B,A{]}=C$ is tantamount to $c<a+b$\,. However, from  (\ref{eq:rhsGradedDegTerms}), a plethora of candidate monomial rearrangement terms exists, at least consistent with graded degree (and thus the resolution of ordering ambiguities), subject to additional inequalities on the parameters $a,b$ and $c$\, which, at least up to cubic degree, do not involve stringent restrictions. We thus consider the following most general degree-three case of commutator-bracket relations, under a suitable assignment of $a,b,c$:
\begin{equation}
\label{eq:generalcubic}
\begin{aligned}
{}[B,A]= &\, a_{300} A^3 + a_{200}  A^2 + a_{100} A +  a_{010} B + a_{001} C \,, \\
[C,A]= &\, b_{300} A^3  +  b_{210}  A^2 B
 +  b_{120} A B^2   +  b_{030} B^3 + 
b_{200} A^2+{}\\ 
 &\, b_{110}  AB  +  b_{020} B^2 + b_{100} A +  b_{010} B + b_{001} C\,,\\
 [C,B]=  &\, c_{300} A^3  +  c_{210}  A^2 B
 +  c_{120} A B^2   +  c_{030} B^3 + c_{200} A^2  + c_{110}  A B  +{}\\
 &\,   c_{101}  A C  +  c_{020} B^2 +
 c_{100} A +  c_{010} B + c_{001} C\,,
\end{aligned}
\end{equation}
for the 26 arbitrary coefficients $a_{001}, a_{010}, \cdots, c_{300}$\,.

As elaborated above, the diamond lemma checks the consistency of these commutator bracket-type relations, as means to define an ordered basis for the enveloping algebra. In this case (see \ref{eq:diamond_CBA}) only the misordered monomial $CBA$ leads to distinct rearrangement step paths, which must stabilize, provided polynomial constraints, in this case of up to quartic degree, are satisfied by the coefficients. Without further assumptions or transformations, however, closed-form solutions are not available. Instead, in order to illustrate the more general approach of the following section, we restrict the discussion here to a limited analysis of admissible extensions to (\ref{eq:daskaloyannisquadratic}), by including terms which incrementally generalize the starting form (\ref{eq:cubic}). For the moment we drop the central terms, and we specialize by setting the free numerical parameters $\alpha, \beta, \cdots$ in (\ref{eq:cubic}) above to unity.

The first case beyond (\ref{eq:cubic}) arises with the addition of a term proportional to $B^2$ in ${[}C,A{]}$. Application of the diamond lemma constrains the additional coefficient, yielding the relations:
\begin{equation}
  \begin{aligned}
    {[}B,A{]}= &\, C\,,\\
    {[}C,A{]}=&\, A^2 + 2 AB + \omega B^2 +A+B+C\,\\
    {[}C,B{]} =&\, A^3 +A^2 -2AB -(1-\omega)B^2 +A+B-C\,;
\end{aligned}  
\end{equation}
thus, as noted above, recovering a member of the Calabi-Yau family \cite{vin20}. 
A further extension is discovered via the introduction of a term proportional to $A^2$ in the ${[}B,A{]}$ bracket, and indeed this modification also admits a cubic extension $A^3$ in ${[}C,A{]}$:
\begin{equation}
\begin{aligned}
    {[}B,A{]}= &\, \rho A^2 + C\,,\\
    {[}C,A{]}=&\, \sigma A^3 + A^2 + 2 AB + \omega B^2 +A+B+C\,\\
    {[}C,B{]} =&\, A^3 -3 \sigma A^2C +A^2 -2AB +(2\rho-3\sigma) AC - {}\\ &\,(1+\sigma\omega-\omega)B^2 +A+(1-\sigma)B-C\,.
\end{aligned}
\end{equation}
Importantly, this algebra cannot be transformed to a known case via a linear basis transformation, and non-linear transformations such as $C'=C+ \rho A^2$, which reduce the order of the ${[}B,A{]}$ bracket to a linear expression, come at the cost of introducing fourth-degree terms into the remaining relations. Reinstating the structure constants as free parameters, we finally present a further family of algebras which in turn generalize the original cubic Daskaloyannis extension \cite{mar09} and the Calabi-Yau forms:
\begin{equation}
   \begin{aligned}
   {}[B,A]= &\,C + a_{200} A^2 \\
[C,A]= &\,b_{300}A^3 +b_{200}A^2 +2 b_{001} A  B+ b_{100} A+b_{020} B^2 + b_{010} B + b_{001} C \\
{}[C,B]= &\,c_{300}A^3 -3 b_{300}A^2 B+c_{200}A^2 -2 (b_{300} b_{001}-2 b_{001} a_{200} +b_{200}) AB +{}\\
&\, (2 a_{200} - 3 b_{300} ) A C +(- b_{001}+ a_{200} b_{020}- b_{020} b_{300}) B^2+ c_{100} A +{}\\
&\, (-b_{300} b_{010}+a_{200} b_{010}-b_{100}) B+(-b_{300} b_{001}+2 a_{200} b_{001}-b_{200}) C
\end{aligned}  
\end{equation}
With the PBW basis in hand, we take a constructive approach to compute the form of a quartic Casimir operator: a polynomial $K$ of fourth degree in the generators $\{A,B,C\}$ such that $[K,A]=[K,B]=[K,C]=0$ (see section \ref{sec:casimir} for details). The structural restrictions imposed by the graded-degree considerations mean that the generator $C$ can only appear as a linear term in the defining relation of the bracket $[A,B]$. As a consequence, it is sufficient to verify $[A,K]=[B,K]=0$ to ensure that $[C,K]=0$. 
\begin{align*}
 K ={}&  3 c_{300} A^4-12 b_{300} A^3B+\big(4 c_{200}+(3 b_{020} (4 a_{200}-3 b_{300})-2 b_{001}\big) c_{300}) A^3+\\
 & 6\big(-2 b_{200}+4 a_{200} (b_{001}-b_{020} b_{300})+b_{020} (6 b_{300}^2+c_{300})\big) A^2B+\\
 & 6(2 a_{200}-3 b_{300}) A^2C+12\big(b_{020} (a_{200}-2 b_{300})-b_{001}\big) AB^2
 -4 b_{020} B^3 + \\
& \big(-c_{300} b_{001}^2-(4 c_{200}+a_{200} b_{020} c_{300}) b_{001}+6 c_{100}+ 8 a_{200} b_{020} c_{200}-\\
&\,\,12 b_{020} b_{300} c_{200}+3 b_{010} c_{300}+3 b_{020} b_{200} c_{300}\big) A^2+\\
& 2 \big(12 b_{001} b_{020} a_{200}^2-6 (2 b_{001}^2+5 b_{020} b_{300} b_{001}-b_{010}+b_{020} b_{200}) a_{200}+12 b_{001} b_{020} b_{300}^2- 6 b_{100}+ \\
&\quad 6 b_{001} b_{200}+6 b_{001}^2 b_{300}-9 b_{010} b_{300}+12 b_{020} b_{200} b_{300}+2 b_{020} c_{200}+2 b_{001} b_{020} c_{300}\big) AB+\\
& 6 \big(2 b_{020} a_{200}^2+(2 b_{001}-7 b_{020} b_{300}) a_{200}+6 b_{020} b_{300}^2-2 b_{200}+b_{001} b_{300}+b_{020} c_{300}\big) AC + \\
& 2 \big(6 b_{001}^2-2 b_{020} (a_{200}-4 b_{300}) b_{001}-3 b_{010}+ \\
&\quad b_{020} (4 b_{020} a_{200}^2-10 b_{020} b_{300} a_{200}-4 b_{200}+b_{020} (6 b_{300}^2+c_{300}))\big)B^2+\\
& 6\big(b_{020} (2 a_{200}-3 b_{300})-2 b_{001}\big) BC+6 C^2+\\
& \big(6 a_{200} b_{020} c_{100}-12 b_{020} b_{300} c_{100}+2 b_{010} c_{200}+3 b_{020} b_{100} c_{300}-b_{001} (6 c_{100}+b_{010} c_{300})\big) A +\\
& 2\big(4 b_{010} b_{020} a_{200}^2-2 b_{020} (2 b_{100}+5 b_{010} b_{300}) a_{200}+6 b_{010} b_{020} b_{300}^2-3 b_{010} b_{200}+ \\
&\quad6 b_{020} b_{100} b_{300}+3 b_{001} (2 b_{100}+b_{010} b_{300})+b_{020} c_{100}+b_{010} b_{020} c_{300}\big)B+\\
& 2\big(6 b_{001} b_{020} a_{200}^2-3 (2 b_{001}^2+5 b_{020} b_{300} b_{001}-b_{010}+b_{020} b_{200}) a_{200}+6 b_{001} b_{020} b_{300}^2-\\
&\quad3 b_{100}+3 b_{001} b_{200}+3 b_{001}^2 b_{300}-3 b_{010} b_{300}+6 b_{020} b_{200} b_{300}+b_{020} c_{200}+b_{001} b_{020} c_{300}\big)C
\end{align*}

We have presented above an incremental series of generalisations of the original three-generator quadratic algebra (\ref{eq:daskaloyannisquadratic}). These generalisations illustrate the potential for the discovery of new algebras, distinct from known cases, thus motivating our development in the following section of a more comprehensive strategy for characterising solutions of the PBW constraints.

\section{Three-generator quadratic PBW-algebras}
\label{sec:Classification}
In this section we provide a detailed study of the class of quadratic commutator PBW-algebras with three generators. Going beyond the investigation of incremental extensions to known PBW algebras (see section \ref{subsec:Deformation}), here we undertake a more complete treatment of the quadratic case to consider all possible solutions to the PBW constraints. Although the application of the diamond lemma to determine constraints on the structural parameters is conceptually straightforward (i.e., the iterated application of the substitution rules), it is in practice, even for just three generators of maximal degree two and using a computer algebra package, a challenging task to characterise the solution space of the constraints: a system of non-homogeneous algebraic equations. The computational challenges include: singularities and the branching of solution paths, intractable numerical algorithms, and the determination of solution uniqueness. To ameliorate these problems, we use a degree-preserving, lower-triangular basis transformation that permits the solution space to be broken into discrete types and sub-types before the application of the diamond lemma. Further branching of the solution types based on the `skew' of the exchange relation (related to $\lambda_{BA}$ in \eqref{eq:diamond_CBA}) and the existence of a cubic Casimir aid both the computational process and presentation of the results.

\subsection{Solution classes via canonical forms}
\label{subsec:solution_classes}
We consider the following most general quadratic (commutator-type) algebra which includes all eligible candidate monomial terms under the natural and most flexible assumption that the graded-degree assignment satisfies all of the inequalities arising in \eqref{eq:rhsGradedDegTerms}:
\begin{equation}
\label{eq:general_quadraticABC}
    \begin{aligned}
{}[B,A] ={}& a_{200} A^2 + a_{110} AB + a_{100} A + a_{010} B + a_{001} C  \\
{}[C,A] ={}& b_{200} A^2 + b_{110} A B  + b_{101} A C + b_{020} B^2 + b_{100} A + b_{010} B + b_{001} C \\
{}[C,B] ={}&  c_{200} A^2 + c_{110} A B  + c_{101} A C  + c_{020} B^2  + c_{011} BC + c_{100}  A  +  c_{010} B  +  c_{001} C 
\end{aligned}
\end{equation}
To simplify the solution space of PBW constraints, we consider the following graded degree-preserving lower-triangular basis transformation 
\begin{equation}
\label{eq:lower_triangular}
    \begin{aligned}
 A' ={}& \alpha_1 A \\
 B' ={}& \beta_1 B + \beta_2 A \\
 C' ={}& \gamma_1 C  + \gamma_2 B + \gamma_3 A
\end{aligned}
\end{equation}
The additional parametric freedom introduced by the basis transformation allows us to `fix' the structure of the commutator $[A',B']$ to one of several canonical forms, providing a useful characterisation of distinct solution classes and significantly simplifying the form of the algebraic equations obtained from the diamond lemma. The canonical forms are those for which the coefficients of all the candidate monomial terms in the defining relation for the bracket $[A',B']$, with exception of the `skew'-term $AB$, are either zero or unity. Taking the bracket-relation of the two transformed generators,
\begin{equation}\label{eq:ABprimeBracket}
\begin{aligned}
 {}[B',A'] ={}& \left(\frac{a_{200}}{\alpha_1^2}- \frac{a_{110} \beta_2}{\alpha_1^2 \beta_1}\right) A'^2 + \frac{a_{110}}{\alpha_1 \beta_1} A'B' + \\
  &\quad\left(\frac{a_{100}}{\alpha_1} - \frac{a_{010} \beta_2}{\alpha_1 \beta_1 } + \frac{ a_{100}\beta_2 \gamma_2}{\alpha_1 \beta_1 \gamma_1 } - \frac{a_{001} \gamma_3}{\alpha_1 \gamma_1}\right) A' + \\
  &\quad \left(\frac{a_{010}}{\beta_1} - \frac{a_{010}\gamma_2 }{\beta_1 \gamma_1} \right)  B' + \frac{a_{001}}{\gamma_1}C'
 \end{aligned}
\end{equation}
the set of canonical forms can be determined by a systematic examination of the allowable combinations, given the dependencies among the monomial coefficients. In practice, this is facilitated by considering separately the three cases: $a_{110}=-1$, $a_{110}=0$ and $a_{110} \neq 0$ (although it turns out that the final list of canonical forms for these three cases coincide). For example, up to the lower-triangular basis transformation, there are only four inequivalent canonical forms containing the term $C'$:\\

\begin{equation}\label{eq:type1terms}
    \begin{aligned}
    &C' + \frac{a_{200}}{\alpha_1 \beta_2} A'B',\\[1mm] 
    &C'  +A'^2 - \frac{\alpha_1}{\beta_2} A'B',\\[1mm]
    &C'+B' + \frac{a_{110}}{\alpha_1\beta_1}A'B', \\[1mm]
    &C'+A'^2 + B' + \frac{a_{110}}{\alpha_1 \beta_1}A'B'.
    \end{aligned}
\end{equation}
In addition to these four forms, there are eight further forms that do not include $C'$. Finally, for the purpose of simplifying the form of the structural PBW constraints, we drop the prime notation to determine the following 12 canonical forms: 
\begin{align*}
    \begin{array}{lccl}
    (1\mathrm{a})\qquad{}& [B,A] &=&  \lambda AB + C\\[0.5mm]
    (1\mathrm{b}) & [B,A] &=& \lambda AB +A^2+ C \\[0.5mm]
    (1\mathrm{c}) & [B,A] &=& \lambda AB +C+B \\[0.5mm]
    (1\mathrm{d}) & [B,A] &=& A^2 + \lambda AB +B+C\\[2mm]
    (2\mathrm{a}) & [B,A] &=& \lambda AB +B\\[0.5mm]
    (2\mathrm{b}) & [B,A] &=&  A^2+ \lambda AB + B \\[0.5mm]
    (2\mathrm{c}) & [B,A] &=& \lambda AB +A\\[0.5mm]
    (2\mathrm{d}) & [B,A] &=& \lambda AB +A +B\\[0.5mm]
    (2\mathrm{e}) & [B,A] &=& A^2 + \lambda AB +A \\[0.5mm]
    (2\mathrm{f}) & [B,A] &=& \lambda AB\\[0.5mm]
    (2\mathrm{g}) & [B,A] &=& A^2 + \lambda AB + B\\[0.5mm]
    (2\mathrm{h}) & [B,A] &=& A^2 + \lambda AB\,, \\[0.5mm]
    \end{array}
\end{align*}
where $\lambda = a_{110}$ and we have chosen to explicitly include the $AB$ term on the right-hand side, rather than including it in the definition of a `skew' bracket, such as $[B,A]_{\lambda_{BA}} := BA - \lambda_{BA} AB$ for $\lambda_{BA} = 1 + a_{110}$. We note that the forms 2(a), ..., 2(h) themselves address the preliminary problem of enumerating canonical forms for quadratic extensions in the two dimensional case.

Here we have applied the lower-triangular basis transformation with the goal of simplification (to aid computation), but the transformation also provides a way to account for the equivalence of quadratic PBW algebras up to rescaling and graded degree-preserving linear basis transformations. To fully account for this type of equivalence would require further characterisation of canonical forms within the remaining bracket relations, exhausting the degrees of freedom in the parameterised transformation. However, even this extended specification of the canonical forms would constitute only a limited form of equivalence, since more complex transformations are available; for example, non-linear transformations or block-diagonal matrix transformations (with added lower-triangular components) with the block structure based on grouped assignment of graded-degree values (i.e., a block corresponds to a group of generators sharing a common graded degree). Given the difficulties of characterising equivalence, we use linear basis transformations for practical computational reasons only and do not attempt to resolve the challenging problem of solution uniqueness for specific instances within each canonical type.

\subsection{Example solutions: form (1a)}
\label{subsec:Exxform1a}
The case (1a) is particularly interesting as it contains the well-known Daskaloyannis and Racah algebras that have proven useful in the study of quantum systems in context of superintegrability \cite{das01,Quesne2007a,Quesne2007b,mil13b}. Here the scope is broader as we are not constrained by requirement of differential realisations or the conservation of a Hermitian quantum-mechanical Hamiltonian. By allowing the skew of the defining relation to vary (i.e., $a_{110}$, $b_{101}$, and $c_{011}$ in \eqref{eq:general_quadraticABC}), and in particular the case that one or more of $a_{110}$, $b_{101}$, and $c_{011}$ equals $-1$, we obtain solution types with similarities to the Leonard pair which do not require relations to be expressed as typical `exchange'-type relations \cite{Terwilliger2005,Terwilliger2004}.

We implement the diamond lemma and solve the corresponding set of constraint equations using the Mathematica package \texttt{NCAlgebra} \cite{ncalgebra} which has the capacity to manipulate noncommutative algebraic expressions. The strategy, from a computational point of view, is to first define the given quadratic, cubic or higher-order polynomial algebra by expressing the relations as a set of substitution rules in terms of the ordered monomials. Next, the diamond lemma is implemented as an iterated (i.e., nested) application of the substitution rules, starting from both the left- and right-hand side of each ambiguous degree-three monomial, to derive the PBW-constraint expressions (i.e., \eqref{eq:diamond_CBA} for the three-generator case). The package functions are used to collect the coefficients of each ordered monomial term, the number of which grows rapidly with the number of generators and the specified maximal polynomial degree of the relations---the computer package can accommodate several thousand terms.

%
%
\newcolumntype{F}{>{\small$}m{10mm}<{$}} 
\newcolumntype{G}{>{\small$}m{120mm}<{$}} 
\renewcommand{\arraystretch}{1.2}
\begin{table}[t!]
\captionsetup{width=0.9\textwidth}
    \centering
\begin{tabular}{F G}
\toprule[0.5mm]
\textbf{term} & \textbf{coefficient} \\
\midrule
 A &  b_{100} b_{200} +  b_{100} c_{001} +  b_{100} b_{200} c_{011} -  b_{001} c_{100} - a_{110}  b_{001} c_{100} +  b_{110} c_{100} +  b_{110} c_{011} c_{100}+  c_{020} c_{100} + a_{110}  c_{020} c_{100} \\
 B &   b_{010} b_{200}  + b_{010} c_{001}  - b_{001} c_{010}  - a_{110} b_{001}  c_{010} + b_{110}  c_{010}  + b_{010} b_{200}  c_{011} + b_{110}  c_{010} c_{011} +  c_{010} c_{020} + a_{110} c_{010} c_{020}\\
 C &  b_{100} + b_{001} b_{200}  - a_{110} b_{001} c_{001}  + b_{110} c_{001} + c_{010}  + b_{100} c_{011}  + b_{001} b_{200} c_{011} + b_{110} c_{001} c_{011}  + c_{001} c_{020}  + a_{110} c_{001} c_{020} \\
 A^2 & b_{200}^2  + b_{200} c_{001} + b_{200}^2 c_{011} - a_{110} c_{100} - b_{101} c_{100} - a_{110} b_{101} c_{100}+ b_{100} c_{101}  - b_{001} c_{200}  - a_{110} b_{001} c_{200}  + b_{110} c_{200} + b_{110} c_{011} c_{200}   + c_{020} c_{200}  + a_{110} c_{020} c_{200} \\
 AB &  b_{110} b_{200} + b_{110} c_{001} - b_{101} c_{010}   - a_{110} b_{101} c_{010}  + b_{100} c_{011}  + a_{110} b_{100} c_{011} + b_{110} b_{200} c_{011}  + b_{010} c_{101} - b_{001} c_{110}  - a_{110} b_{001} c_{110}  + b_{110} c_{110}  + b_{110} c_{011} c_{110}+ c_{020} c_{110}  + a_{110} c_{020} c_{110}\\
 AC & 2 b_{200}  + a_{110} b_{200}  + b_{101} b_{200}  - a_{110} c_{001} - a_{110} b_{101} c_{001} + 2 b_{200} c_{011}+ a_{110} b_{200} c_{011} + b_{101} b_{200} c_{011} - a_{110} b_{001} c_{101} + b_{110} c_{101} + b_{110} c_{011} c_{101}+ c_{020} c_{101}  + a_{110} c_{020} c_{101}  + c_{110} \\
 B^2 & -a_{110} b_{010} + b_{020} b_{200} + b_{020}c_{001} + b_{010} c_{011} + b_{020} b_{200} c_{011} -b_{001} c_{020} - a_{110} b_{001} c_{020} + c_{001} + b_{010} c_{011}  + b_{020} b_{200} c_{011} - b_{001} c_{020} - a_{110} b_{001} c_{020} + b_{110} c_{020} + b_{110} c_{011} c_{020}  + c_{020}^2  + a_{110} c_{020}^2\\
 BC & - a_{110} b_{001}  + b_{110} - a_{110} b_{001} c_{011}  + 2 b_{110} c_{011}  + b_{110} c_{011}^2  + 2 c_{020}  + a_{110} c_{020}+ c_{011} c_{020} + a_{110} c_{011} c_{020} \\
 C^2 & b_{101}  + c_{011}  + b_{101} c_{011} \\
 A^3 & b_{200} c_{101}  - a_{110} c_{200}  - b_{101} c_{200}  - a_{110} b_{101} c_{200} \\
A^2B & a_{110} b_{200} + a_{110}^2 b_{200} + b_{200} c_{011} + 2 a_{110} b_{200} c_{011} + a_{110}^2 b_{200} c_{011} + b_{110} c_{101}  - b_{101} c_{110}  - a_{110} b_{101} c_{110}\\
 A^2C & - a_{110} c_{101}  - a_{110} b_{101} c_{101} \\
 AB^2 & b_{110} c_{011}  + a_{110} b_{110} c_{011}  + a_{110} c_{020}  + a_{110}^2 c_{020}  - b_{101} c_{020}  - a_{110} b_{101} c_{020} + b_{020} c_{101} \\
 B^3 & - a_{110} b_{020}  + b_{020} c_{011}\\
 \bottomrule[0.5mm]
\end{tabular}
    \caption{PBW constraints for the general three-generator quadratic algebra. The polynomial expressions in the coefficient column must be simultaneously zero to satisfy the generalised Jacobi identities arising from the diamond lemma.}
    \label{tab:pbw_constraints}
\end{table}
For the ordered monomials to form a PBW basis, the diamond lemma demands that each monomial term vanishes simultaneously, yielding a set of algebraic constraints on the structure constants. For the (1a) case, the independent monomials are the ordered terms $A$, $B$, $C$, $AB$, $AC$, $BC$,$A^2$,$B^2$, $C^2$, $A^3$, $A^2 B$, $A B^2$, and $B^3$, for which the associated algebraic systems of equations have at most cubic degree. The complete set of equations is given in Table \ref{tab:pbw_constraints}. We provide below some example solutions for the canonical forms (1a) differentiated by the existence of a cubic Casimir (see section \ref{sec:casimir} for details) and the allowable values of the parameter $\lambda = a_{110}$.

\subsubsection*{Solution with a cubic Casimir}
\noindent For the canonical form (1a), there is a unique parametric solution that possesses a cubic Casimir:
\vspace{0mm}

\noindent${}\quad\bullet\,$  Form (1a): $\lambda\neq-1$
\begin{align}\label{eq:sol_1a_casimir}
    \begin{array}{llcl}
    &{}[B,A] &=&  C  + \lambda AB\\[1mm]
    &{}[C,A] &=&      \alpha A^2 - \frac{\lambda +2}{\lambda+1}\beta AB   -\frac{\lambda}{\lambda+1}AC + {}\\[1mm]
    &&& \quad \gamma B^2+\frac{1}{\lambda+1}\delta A + \epsilon B - \frac{1}{\lambda+1}\beta C\\[1mm]
    &{}[C,B] &=&    \zeta A^2  -  (\lambda+1)(\lambda+2)\alpha AB +  \beta B^2 +{}\\[1mm]
    &&&\quad  \lambda BC  + \eta A  + \delta B - (\lambda+ 1)\alpha C
    \end{array}
\end{align}
This family of solutions generalises the Racah and Daskaloyannis algebras \eqref{eq:daskaloyannisquadratic}. The solution constraints on the original coefficients $a,b,c$ are easily identified by comparing \eqref{eq:sol_1a_casimir} with \eqref{eq:general_quadraticABC}; for example, $\lambda = a_{110}$, and $\beta = -\frac{\lambda+1}{\lambda+2}b_{110}=-(\lambda+1)b_{001} = c_{020}$. The form of the cubic Casimir is presented in section \ref{sec:casimir}. 

\subsubsection*{Examples of solutions without a Casimir}
\noindent In contrast to the single solution family above, there are many distinct solutions that do not possess a Casimir invariant. We provide three examples below:
\vspace{5mm}

\noindent$\quad\bullet\,$ Form (1a): $\lambda \neq 0,-1$
\begin{align}\label{eq:sol_1a_nocasimir_lambdaneqzero}
\begin{aligned}
 {}[B,A] ={}& \lambda A B + C \\
 {}[C,A] ={}&  \lambda A C - \tfrac{ (1+\lambda)^2 \alpha}{\lambda^2} B + \tfrac{2\alpha(1 + \lambda)}{\lambda} C    \\
 {}[C,B] ={}&  \alpha B^2  - \tfrac{\lambda}{1+\lambda} BC 
 \end{aligned}
\end{align}
\vspace{2mm}

\noindent$\quad\bullet\,$  Form (1a): $\lambda = 0$ ($\alpha\neq 0$); the generators $A$ and $C$ form a quadratic subalgebra
\begin{align}\label{eq:sol_1a_nocasimir_lambdaeqzero}
\begin{aligned}
 {}[B,A] ={}& C \\
 {}[C,A] ={}&  -\tfrac{\alpha}{1+\alpha} AC  + \beta A   \\
 {}[C,B] ={}& \gamma AC + \alpha BC - \tfrac{(1 +  \alpha)\beta \gamma}{\alpha} A -
 (1 +  \alpha)\beta B
 \end{aligned}
\end{align}
\vspace{2mm}

\noindent$\quad\bullet\,$  Form (1a): $\lambda = -1$ ($\gamma\neq -1$ and $\beta \neq 0$)
\begin{align}\label{eq:sol_1a_neqmin1}
\begin{aligned}
 {}[B,A]={}& -AB+C \\
 [C,A]={}&  \alpha A^2 + \beta A B  - \tfrac{\gamma}{1 + \gamma} AC  + \alpha\beta (1 + \gamma) A  \\
 [C,B]={}& \tfrac{-\alpha - \delta - \alpha \gamma}{1 + \gamma}AB + \epsilon B^2 + \gamma BC -{} \\ 
 & \quad \alpha ( \alpha + \delta + \alpha \gamma)  A  -  \tfrac{\zeta (\alpha + \delta + \alpha \gamma)}{\beta(1 +\gamma)} B +  \delta C
\end{aligned}
\end{align}

\noindent We note as a general feature of the (1a) form that all solutions for the case $\lambda=-1$ fail to admit a Casimir invariant. Indeed, the $\lambda=-1$ solutions comprise a special subclass of quadratic algebras as they depart from the usual `skew-bracket' scenario where a misordered pair is brought into fully reduced form by simply exchanging the order of the generators in the pair, up to an overall scaling and the possible addition of lower degree terms. In the $\lambda=-1$ case, the misordered pair $BA$ is replaced with an ordered polynomial expression that may not even contain the original generators (e.g., $BA \mapsto C$ in  \eqref{eq:sol_1a_neqmin1}), while the remaining substitution rules retain the usual form. The possible mixing of these two types of defining relations is a general feature of the class of PBW algebras. 

Further special forms captured by the solutions are evident from examination
of various special cases. For example in (\ref{eq:sol_1a_casimir}) above, the choice $\lambda=-2$ 
turns all bracket relations into anticommutators, yielding a quadratic extension of 
the symmetric three element Jordan algebra with defining relations $\{A,B\}=C$\,, and cyclic permutations
 (c.f. \cite{BrackenGreen1973}). For the choice $\lambda=-2$ of 
(\ref{eq:sol_1a_nocasimir_lambdaneqzero})
however, $B^2$ replaces $A$ in the $\{B,C\}$ anticommutator at the expense of a combination of both $B$ and $C$ in the $\{C, A\}$ relation.
On the other hand, the case $\alpha=-2$ of (\ref{eq:sol_1a_nocasimir_lambdaeqzero}) admits
a ${\mathbb Z}_2$ grading, with $A,B$ odd and $C$ even, but with  (anti)commutation relations `opposite' to those in a Lie superalgebra, in that the odd generators have a commutator bracket, while anticommutator brackets occur between odd and even generators. Features such as the persistence of anticommutator brackets for some parameter choices, also recur in the remaining cases beyond the (1a) canonical form, to which we now turn.
\subsection{Solutions for other canonical forms}
\label{subsec:solns_other}
\noindent Solutions for the canonical forms (1b) to (2h) can be obtained in the same way as we have demonstrated above for the form (1a). In each case, the constraints arising from the diamond lemma lead to 
new quadratic PBW algebras, both with and without cubic Casimirs for various constraint conditions on the parameter $\lambda = a_{110}$. To illustrate some different solution types, we present here two examples 
which do admit a cubic Casimir invariant (to be presented explicitly in section \ref{sec:casimir}):
\vspace{5mm}

\noindent$\quad\bullet\,$ Form (1b): $\lambda=0$ 
\begin{align}\label{eq_qa_soln_1b}
\begin{aligned}
 {}[B,A]={}& A^2 + C\\
 [C,A]={}& \alpha A^2 + \beta A B  + \gamma B^2 +  \delta A + \epsilon B + \tfrac{\beta}{2}  C \\
 [C,B]={}&  \zeta A^2 +  2 (\beta - \alpha) A B  + 2 A C  + \tfrac{1}{2} (2 \gamma - \beta) B^2  + {}\\
 & \quad \eta  A  +  (\epsilon-\delta) B  +  ( \beta - \alpha ) C 
 \end{aligned}
\end{align}
\vspace{2mm}
$\quad\bullet\,$ Form (2d): $\lambda\neq -1$ 
\begin{align}\label{eq_qa_soln_2d}
\begin{aligned}
 {}[B,A]={}& \lambda AB  + A + B \\
 [C,A]={}& \alpha A^2 - \tfrac{2 \beta + \lambda}{1 + \lambda}\beta AB - \tfrac{\lambda}{1 + \lambda} AC +{}  \\
 & \quad \gamma A + \delta B - \tfrac{1}{1 + \lambda} C\\
 [C,B]={}&  \epsilon A^2 - (\lambda+1)(\lambda+2) \alpha AB  + \beta B^2  + \lambda B C \\
 & \quad \zeta  A  - ((\lambda+1)(\gamma+\alpha)+\beta) B  +  C
 \end{aligned}
\end{align}

We have introduced and illustrated with several example cases an algorithmic method to explore the solution space of three-generator quadratic PBW algebras. As emphasized in section \ref{subsec:solution_classes}, due to the difficulty of defining solution equivalence, the introduced methods exhaust all possible solutions but do not lead to a unique characterisation of solution independence. 

\section{Casimir Invariants}
\label{sec:casimir}
\noindent
In the case of Lie algebras, the study of Casimir invariants is important; in particular, due to the connection with representation theory, the possibility of using Casimirs in state labelling, or even deriving results for physical spectra or other data. While general formulae based on the Killing form are known for simple and semi-simple Lie algebras \cite{per65, per68,rac65,pau66}, the situation for non semisimple Lie algebras is more difficult. Often different methods have been proposed on a case-by-case basis, each of them with specific limitations. Some explicit formulae are known for low dimensional algebras (for example, up to dimension six  \cite{sno14,bel66,cam05}\,), and for some families of Lie algebras such as the Schr\"{o}dinger algebra. 

The case of quadratic or even polynomial algebras in a PBW basis is very similar to that of the non-semisimple Lie algebras. Here general formulae based on structural properties are generally unavailable, and Casimir invariants do not necessarily exist, depending on the given set of structure constants. One possible approach used for Lie algebras is to investigate the problem in the associated classical Poisson-Lie algebra via the Kirillov correspondence \cite{sno14,Kir1976}. Instead of this, here we use symbolic computation and the \texttt{NCalgebra} package \cite{ncalgebra} to address the problem directly in the commutator algebra setting.

We therefore adopt the following definition of a Casimir invariant, expressed here for the three-generator case, which is constructed as a degree $n$ polynomial in the enveloping algebra in the PBW basis,
\begin{align}\label{eq_define_Casimir}
K= \sum_{n_1\!+\!n_2\!+\!n_3 \leq n} \gamma_{n_1,n_2,n_3} A^{n_1} B^{n_2} C^{n_3} 
\end{align}
such that 
\begin{align}\label{eq_K_commutators}
    [K,A]=[K,B]=[K,C]=0.
\end{align}
The Casimir invariant will be said to be of order $n$ where $n$ is the polynomial degree of $K$ \eqref{eq_define_Casimir}. The $\gamma_{n_1,n_2,n_3}$ are then to be given in terms of the structure constants of the quadratic algebra. 

Based on the example solutions provided in section \ref{subsec:Exxform1a}, wherein the quadratic algebras admitted invariants of degree 3, we here restrict attention to cubic degree. Explicitly, we construct the Casimir as $K=K^{(1)} + K^{(2)} + K^{(3)}$, with
\begin{align}
    K^{(i)}=  \sum_{n_1 \!+\!n_2 \!+\!n_3 =i} \hskip-3ex \gamma^{(i)}_{n_1,n_2,n_3} A^{n_1} B^{n_2} C^{n_3} \,,\quad
i=1,2,3
\,,
\end{align}
for which satisfaction of \eqref{eq_K_commutators} leads to a linear system of equations for the given parameters $\gamma^{(i)}_{n_1,n_2,n_3}$, $i=1,2,3$. %

The solutions of these constraint equations will be the linearly independent Casimir invariant operators. Finally, one needs to look for any algebraic relations amongst such candidates. Here we will be looking for invariants which satisfy the constraints, without imposing further conditions on the structure constants.

\subsection{Explicit form of the cubic Casimir for form (1a)}
There is a unique solution possessing a cubic Casimir:

\noindent$\quad\bullet\,$ Form (1a) [Solution \eqref{eq:sol_1a_casimir}, $\lambda \neq -1$]
\begin{align}\label{eq_Casimir_1a}
    \begin{aligned}
     K &=   -\left(\beta  \eta  \lambda ^2+3 \beta  \eta  \lambda +3 \beta  \eta +\zeta  \lambda  \epsilon +\zeta  \epsilon \right)  (\lambda +1) A -{} \\
     &\quad \left(-\alpha  \lambda ^4 \epsilon -5 \alpha  \lambda ^3 \epsilon -10 \alpha  \lambda ^2 \epsilon -9 \alpha  \lambda  \epsilon -3 \alpha  \epsilon +\beta  \delta  \lambda ^3+ {} \right.\\
     &\left.\qquad 5 \beta  \delta  \lambda ^2+9 \beta  \delta  \lambda +6 \beta  \delta +\gamma  \eta  \lambda ^2+2 \gamma  \eta  \lambda +\gamma  \eta \right) B +
     \gamma  (\lambda +1) (\lambda +2) B^3 - {} \\ &\quad\left(-\alpha  \beta  \lambda ^3-4 \alpha  \beta  \lambda ^2-6 \alpha  \beta  \lambda -3 \alpha  \beta +\gamma  \zeta  \lambda +\gamma  \zeta +\delta  \lambda ^2+3 \delta  \lambda +3 \delta \right)  (\lambda +1)  C -{} \\
     &\quad \left(\beta  \zeta  \lambda ^2+3 \beta  \zeta  \lambda +2 \beta  \zeta +\eta  \lambda ^3+4 \eta  \lambda ^2+6 \eta  \lambda +3 \eta \right)  (\lambda +1) A^2 -{} \\
     &\quad (\lambda +1) (\lambda +2) \left(-\alpha  \beta  \lambda ^3-4 \alpha  \beta  \lambda ^2-6 \alpha  \beta  \lambda -3 \alpha  \beta + {}\right.\\
     &\qquad \left.\gamma  \zeta  \lambda +\gamma  \zeta +\delta  \lambda ^2+3 \delta  \lambda +3 \delta \right) AB -{} \\
     &\quad \left(-\alpha  \lambda ^3-4 \alpha  \lambda ^2-6 \alpha  \lambda -3 \alpha \right)  (\lambda +1) (\lambda +2)  AC - {} \\
     &\quad \left(-\alpha  \gamma  \lambda ^4-6 \alpha  \gamma  \lambda ^3-13 \alpha  \gamma  \lambda ^2-12 \alpha  \gamma  \lambda -4 \alpha  \gamma +\beta ^2 \lambda ^3+ {} \right. \\ &\qquad \left. 5 \beta ^2 \lambda ^2+9 \beta ^2 \lambda +6 \beta ^2-\lambda ^3 \epsilon -4 \lambda ^2 \epsilon -6 \lambda  \epsilon -3 \epsilon \right) B^2 -{}\\
     &\quad \beta  \left(\lambda ^2+3 \lambda +3\right)  (\lambda +1) (\lambda +2)  BC - \left(\lambda ^2+3 \lambda +3\right) (\lambda +1)^2 C^2 + {} \\
     &\quad \alpha  \left(\lambda ^2+3 \lambda +3\right) (\lambda +1)^3  (\lambda +2)  A^2 B - {} \\
     &\quad \left(\beta  \lambda ^2+3 \beta  \lambda +3 \beta \right)  (\lambda +1) (\lambda +2) AB^2 - {} \\
     &\quad \left(\lambda ^3+3 \lambda ^2+3 \lambda \right)  (\lambda +1) (\lambda +2)  ABC -\zeta  (\lambda +1)^3  (\lambda +2) A^3
    \end{aligned}
\end{align}

\subsection{ Casimirs for some other canonical forms}
We provide below the cubic Casimir invariants for the two quadratic algebra solutions given in section \ref{subsec:solns_other}:
\vspace{3mm}

\noindent$\quad\bullet\,$ Form (1b) [Solution \eqref{eq_qa_soln_1b}, $\lambda = 0$]
\begin{align}
    \begin{aligned}
     K &= -\tfrac{1}{6} (3 \beta  \eta -6 \gamma  \eta -2 \zeta  \epsilon ) A -\tfrac{1}{3}  (3 \alpha  \epsilon -3 \beta  \delta +4 \gamma  \delta -\gamma  \eta -4 \gamma  \epsilon ) B - {} \\
     &\quad  \tfrac{1}{6}  \left(-3 \alpha  \beta +6 \alpha  \gamma +3 \beta ^2-6 \beta  \gamma -2 \gamma  \zeta +6 \delta -6 \epsilon \right) C -{}\\
     &\quad \tfrac{1}{3} (\beta  \zeta -4 \gamma  \zeta -3 \eta ) A^2
 -\tfrac{1}{3}  \left(-3 \alpha  \beta +6 \alpha  \gamma +3 \beta ^2-6 \beta  \gamma -2 \gamma  \zeta +6 \delta -6 \epsilon \right)  AB - {} \\
 &\quad \tfrac{1}{6}  \left(8 \alpha  \gamma -3 \beta ^2+2 \beta  \gamma -8 \gamma ^2+6 \epsilon \right) B^2  -(\beta -2 \gamma ) BC + C^2 + {}\\
 &\quad 2 (\beta -\alpha )  A^2 B - (\beta -2 \gamma ) AB^2- (2 \alpha -\beta -2 \gamma ) AC  +2 A^2 C +\tfrac{2}{3} A^3 \zeta  -\tfrac{2}{3} \gamma  B^3
    \end{aligned}
\end{align}

\vspace{3mm}
\noindent$\quad\bullet\,$ Form (2d) [Solution \eqref{eq_qa_soln_2d}, $\lambda \neq -1$]
\begin{align}
    \begin{aligned}
      K &=\left(\zeta  \lambda ^2+3 \zeta  \lambda +3 \zeta +\epsilon \right) (\lambda +1) A - {}\\
      &\quad \left(\lambda ^2+3 \lambda +3\right) \left((\alpha +\gamma )(\lambda +1) (\lambda +2)  + (\beta  \lambda +2 \beta -\delta  \lambda -\delta )\right) B - {}\\
      &\quad \left(-\zeta  \lambda ^2-3 \zeta  \lambda -3 \zeta -\lambda  \epsilon -3 \epsilon \right)  (\lambda +1)^2 A^2 -{}\\
      &\quad \left(\lambda ^2+3 \lambda +3\right)  (2 \alpha  \lambda +2 \alpha +\beta +\gamma  \lambda +\gamma )  (\lambda +1) (\lambda +2) AB -{}\\
      &\quad \left(-\lambda ^2-3 \lambda -3\right)   (\lambda +1) (\lambda +2) AC- \left(\lambda ^2+3 \lambda +3\right) (\beta  (-\lambda )-2 \beta +\delta  \lambda +\delta ) B^2 -{}\\
      &\quad \left(-\lambda ^2-3 \lambda -3\right)  (\lambda +1) (\lambda +2) BC +{}\\
      &\quad \epsilon  (\lambda +1)^3  (\lambda +2) A^3- \alpha  \left(\lambda ^2+3 \lambda +3\right)   (\lambda +1)^3  (\lambda +2)  A^2 B + {}\\
      &\quad \beta  \left(\lambda ^2+3 \lambda +3\right)   (\lambda +1) (\lambda +2) AB^2 +\lambda  \left(\lambda ^2+3 \lambda +3\right)   (\lambda +1) (\lambda +2) ABC
    \end{aligned}
\end{align}

\section{Conclusions}
In this paper we have presented a class of generalized commutator algebras of PBW type. This class include well-known examples such as the quadratic Racah and Daskaloyannis algebras which have applications in the context of quantum superintegrable systems. We described solution strategies and different properties of this class of quadratic and polynomial algebras more generally . 

We have introduced a general approach for obtaining parametric solutions of the structure constants based on a concrete construction of the PBW constraints via application of the diamond lemma. Using a degree-preserving basis transformation, we showed that the solution space for the class of quadratic commutator PBW algebras can be divided into types 1 and 2, where type 2 permits a two-dimensional subalgebra $\{A,B\}$. These two general types split into four and eight sub-types, or canonical forms, respectively. Solutions within each form can be further organised according to allowable values of the `skew' parameter $\lambda$, which appears in the substitution rule associated with the bracket $[A,B]$, and the existence of a cubic Casimir operator. Applying the general construction, we derived explicit parametric solutions for the canonical form (1a) and for some selected cases within other forms; for each solution we computed the corresponding cubic Casimir, if it existed. 

For the form (1a), there is a unique parametric solution possessing a cubic Casimir---this particular case generalises the quadratic Racah and Daskaloyannis algebras mentioned above. The examples we have provided for (1a) without a Casimir and for (1b) and (2d) with a cubic Casimir, show that the flexibility of the defining relations, for which the general forms are only mildly constrained by graded-degree considerations, coupled with preservation of the usual ordered PBW basis, leads to new and interesting classes of algebras which may be relevant in the context of mathematical physics and the investigation of non-commutative algebraic structures more broadly. 

Quadratic and cubic algebras have already been discovered in context of quantum superintegrable systems. In particular, interesting cubic algebras are related to models associated with the fourth \cite{mar09b} and sixth \cite{mar20} Painlev\'{e} transcendents. These examples also reveal interesting connections with the theory of nonlinear differential equations. Cubic algebras have been connected with the theory of missing labels and Gaudin models \cite{vin20}. These examples demonstrate the importance of such algebraic methods, including structural classification, in mathematical physics.

Although we have focused on the three-generator quadratic case, well-known cubic and higher degree algebras should also find their natural place within parametrised solutions of their corresponding classes of PBW algebras. Our explicit methodology extends naturally to algebras with more than three generators wherein the defining relation associated with the `lowest' bracket (i.e., [B,A] where $d(BA)=\mathrm{min}\{d(x) \mid x\in LM(R)\}$) retains its role as the expression on which structural constraints are imposed to characterise canonical forms. Further investigation of parametrised basis transformations, extending the lower-triangular transformations presented in section \ref{subsec:solution_classes}, could provide insights into the characterisation of solution independence and ultimately a pathway to classification.

Central extensions play a key role in certain Lie algebras; for example, in the definition of the Schr\"{o}dinger algebra and various other conformal algebras with applications in physics. Central extensions are also important in regard to polynomial algebras. Indeed, quantum Hamiltonian approaches lead naturally to central extensions \cite{das01}. To illustrate the potential for central extensions in the current framework, we provide in Appendix \ref{appendixA} one example of a quadratic commutator algebra with a central extension added to each commutation relation; however, due to the complexity of the general problem, a more complete discussion of admissible central extensions for generalized commutator PBW algebras is beyond the scope of the present paper.

An open problem in the context of quantum superintegrable systems is to gain an algebraic understanding of certain aspects of contractions as they are currently applied to realized instances. Contractions are singular basis transformations that can be used to connect Lie algebras that are not isomorphic. Contractions play a key role in describing various non semi-simple Lie algebras. It has been demonstrated using quadratic algebras of superintegrable systems how contraction, based on realizations of Inonu-Wigner type, can be used to relate quantum superintegrable systems to their orthogonal polynomials \cite{mil13b}. A more algebraic understanding of these aspects of contractions, based, for example, on extensions of the present work, could provide further insight and in particular, the possible classification of superintegrable systems.

\subsection*{Acknowledgements:}

IM was supported by by Australian Research Council Future Fellowship FT180100099. 

\subsection*{Data availability statement}

The computer code to reproduce the findings of this study is available upon request from the authors.

\begin{appendix}
\section{Appendix: central extensions}\label{appendixA}
In context of quantum superintegrable systems, the related symmetry algebras (e.g., the Daskaloyannis algebra \eqref{eq:daskaloyannisquadratic}) admit central extensions; this is a general feature of algebraic structures in quantum systems with symmetry algebra. To keep the computational problem tractable, we did not include central extensions for the class of quadratic commutator PBW algebras examined in section \ref{sec:Classification}. Indeed, central extensions are a difficult problem in general for polynomial algebras. Here we present an explicit example which illustrates how the algebraic constraints are modified by the addition of central extensions. Extending \eqref{eq:general_quadraticABC}, we take the following general form:
\begin{align}
    \begin{aligned}
      {}[B,A]&= a_{001} C +  a_{200} A^2  + a_{010} B  + a_{100} A + a_{110} AB + c_1 \\
      [C,A]&= b_{200} A^2 + b_{110} A B  + b_{101} A C + b_{020} B^2 + b_{100} A + b_{010} B + b_{001} C  +  c_2\\
      [C,B]&= c_{200} A^2 + c_{110} A B  + c_{101} A C  + c_{020} B^2  + c_{100}  A  +  c_{010} B  +  c_{001} C  + c_3
    \end{aligned}
\end{align}
The Jacobi identity constraints are derived in the usual way from application of diamond lemma. For the case $\lambda = a_{110}=-1$, the constraints read
\begin{align}
    \begin{aligned}
      0 &= b_{110} (c_{011}+1 ) (b_{200} (c_{011}+1 ) c_1 + c_3 )-\tfrac{b_{10} c_1 (b_{200} c_{011}+b_{200}+c_1 )}{b_{110} (c_{011}+1 )}+ {} \\
      &\qquad c_2 (b_{200} c_{011}+b_{200}+c_{001})\\
      0 &= b_{200} c_{011} c_1 -\tfrac{c_{001} c_1}{c_{011}+1}+ c_3 \\
      0 &= (c_{011}+1) (b_{110} c_1 + c_2 )-\tfrac{b_{010} c_1}{b_{110}}\,, 
    \end{aligned}
\end{align}
with solutions
\begin{align}
    \begin{aligned}
      c_2 &= -\frac{c_1 (b_{110}^2 c_{011}+b_{110}^2-b_{010})}{b_{110} (c_{011}+1)}\\
      c_3 &= c_1 (\frac{c_{001}}{c_{011}+1}-b_{200} c_{011}).
    \end{aligned}
\end{align}
Thus, there are two relations among the three central extensions with only one of them independent. This provides a family of centrally extended algebras. 
\end{appendix}

\bibliographystyle{ieeetr}
\bibliography{refs}

\end{document}